\documentclass[sigconf ]{acmart}

\usepackage{booktabs} 
\usepackage{algorithm,algpseudocode}
\usepackage{graphicx} 
\usepackage{subcaption}
\usepackage[inline]{enumitem}
\usepackage{multirow}

\setcopyright{rightsretained}
\input{Definitions}

\newcommand{\Ebb}{{\mathbb E}}

\acmDOI{}

\acmISBN{}

\acmConference[]{}
\acmYear{}
\copyrightyear{}

\acmArticle{}
\acmPrice{}

\begin{document}
\title{Personalized Treatment Selection using Causal Heterogeneity}

\author{Ye Tu, Kinjal Basu, Cyrus DiCiccio, Romil Bansal, Preetam Nandy, Padmini Jaikumar, Shaunak Chatterjee}
\affiliation{\institution{LinkedIn Corporation}}
\email{{ytu, kbasu, cdiciccio, robansal, pnandy, pjaikumar, shchatterjee}@linkedin.com}







\renewcommand{\shortauthors}{Tu et al.}

\begin{abstract}
Randomized experimentation (also known as A/B testing or bucket testing) is widely used in the internet industry to measure the metric impact obtained by different treatment variants. A/B tests identify the treatment variant showing the best performance, which then becomes the chosen or selected treatment for the entire population. However, the effect of a given treatment can differ across experimental units and a \textbf{personalized approach for treatment selection} can greatly improve upon the usual global selection strategy. In this work, we develop a framework for personalization through (i) estimation of heterogeneous treatment effect at either a cohort or member-level, followed by (ii) selection of optimal treatment variants for cohorts (or members) obtained through (deterministic or stochastic) constrained optimization. 

We perform a two-fold evaluation of our proposed methods. First, a simulation analysis is conducted to study the effect of personalized treatment selection under carefully controlled settings. This simulation illustrates the differences between the proposed methods and the suitability of each with increasing uncertainty. We also demonstrate the effectiveness of the method through a real-life example related to serving notifications at Linkedin. The solution significantly outperformed both heuristic solutions and the global treatment selection baseline leading to a sizable win on top-line metrics like member visits. 
\end{abstract}

%
%



\maketitle

\section{Introduction}
\label{sec:introduction}

In large-scale social media platforms, each member is part of several randomized experiments, also called A/B tests. Their experience is determined collectively by the treatment variants that are selected for them in each of those experiments. Such treatment variants could be different machine learning models, parameter value choices within a composite recommendation system, and UI components (e.g., the font size of specific elements, copy testing). It is common practice to identify the treatment variant that performs the best in the entire population and ramp that variant to everyone. We refer to this practice as ``global allocation''.

Global allocation can be suboptimal. The effect of a treatment variant on individual members (or member cohorts) can be very different. Consider the choice of tone in a marketing email. Younger users of a platform may (on aggregate) prefer an informal tone while older users may prefer something more formal. Global allocation will subject one of these groups to a poorer experience than what may be possible if we selected different copy variants for different member cohorts. Personalized treatment selection can thus enable better member experience and bigger business wins. There is also an important side benefit worth highlighting. As a society, we are today striving to build experiences that are more inclusive. The ability to select treatments in a personalized fashion can be immensely helpful in improving the experience of under-represented groups, especially if they have different preferences. Since the average effect is dominated by the effect on the majority class of users, the global allocation can not only result in loss of business metrics, it can also (unintentionally) make a platform or product less inclusive.

\begin{figure}[!h]
        \centering
        \includegraphics[scale = 0.7]{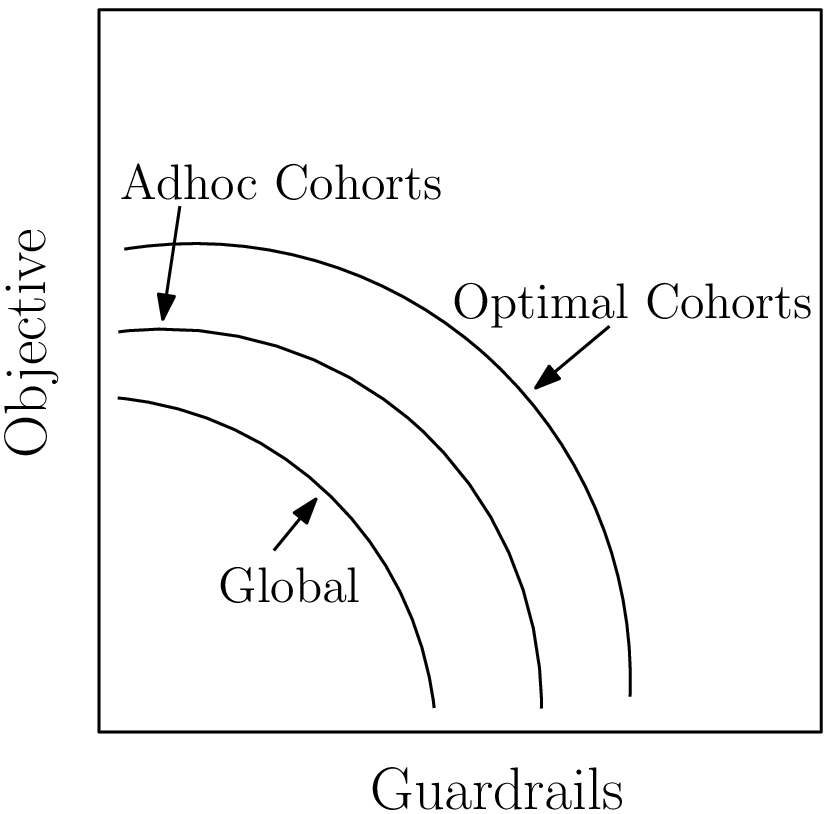}
        \caption{Pareto Optimal Curves for Objective vs Guardrail metrics. Since cohort level selection offers more flexibility than global allocation, the performance envelope becomes strictly better.}
        \label{fig:pareto}
    \end{figure}

Let us consider a general A/B test set up with an objective and a guardrail metric (both at the global population level) as shown in Figure \ref{fig:pareto}. The Pareto frontier is derived by using different choices within the specific family of solutions. The global allocation gives an inferior Pareto optimal curve and choosing variants to allocate to ad-hoc cohorts is strictly better. The difference will depend on the choice of cohorts and hence choosing these cohorts wisely is one of our main focus areas. 

Randomized experiments \cite{kohavi2013online, kohavi2014seven, tang2010overlapping, xu2015infrastructure} can be used to identify the individual preferences by estimating the causal effect of a variant on business metrics which proxy member preference (e.g., total clicks, total sessions, etc.). Traditional A/B testing, however, only gives us the \textit{average} effect of the treatment variant. \textit{Getting a member-level causal effect of a variant is extremely challenging since it is fundamentally unobservable}. However, recent work \cite{Athey7353, stefan2015forest} has shown how we can leverage randomized experiment data to come up with cohorts of members that portray heterogeneity in the causal effects. Identifying this heterogeneity is useful, but does not serve the end goal of improving the member experience and deliver better business objectives. We build on prior work \cite{Athey7353, stefan2015forest, uplift} for cohort identification with a slight extension to handle multiple treatments and multiple metrics. We complement that piece with an optimization formulation that enables us to serve our end goal. Furthermore, since we want our end solution to handle uncertainty well, the variance aware cohort identification methods \cite{Athey7353} are a great and opportunistic fit for our overall methodology.

Throughout this paper, we assume that we have the ability to run randomized experiments on members with different treatment variants, and capture the effect of the treatment on member-related actions (i.e., business metrics) such as clicks, comments, views, scrolls, etc. Ideally, for all personalized recommendations (e.g., which item or ad to show to a user on her feed), we should use causal data. However, the amount of data needed for this is infeasible and hence we rely on observational data and predictive models based on correlation. For personalization of treatment variants, however, using causal data is much more feasible since the number of choices is much smaller. Feasibility is further facilitated by the cohort-level selection, instead of the member-level.

Some applications may have a requirement for global treatment allocation. For instance, when a platform tests the background color of their mobile app, there may be a strong preference to converge to a single new color and form a brand identity around that. Related issues around stability and consistent user experience might also warrant a single global parameter or treatment to be ramped across all users. In such applications, while there may be room for heterogeneous treatments, temporal consistency on a per-member basis may be desired (e.g., change the font size for a given member only once a year). One way to attain any desired temporal consistency is by modulating how frequently the heterogeneous treatment allocation module is run. Finally, in applications where the amount of data from randomized experiments is small, there may not be enough data to estimate any heterogeneity and hence allocating the global best variant is prudent. Our methods also converge to this conclusion in such a scenario.

We conduct simulation analysis to evaluate and compare each of the proposed methods in different scenarios (especially across distinct levels of noise). We discuss how to pick a particular method for an application and show the benefit of the chosen approach (using previously stated guidelines) in a real-world application on Notifications at LinkedIn. Note that our overall framework is general and is applicable to many problems of a similar nature. Both offline simulations and online A/B tests show that our solution performs significantly better than both heuristic solutions as well as global allocation. While causal heterogeneity estimation has been well studied in the recent past, our proposals in this work are the first (to the best of our knowledge) to provide a principled end-to-end solution for identifying and exploiting this heterogeneity to improve member experiences, deliver more business value and build more inclusive products.

The major contributions of our paper are as follows.
\begin{itemize}
\item We develop a general framework of selecting optimal treatment variants for members by estimating heterogeneous causal effects and solving an optimization problem.
\item We discuss ways to identify which among the proposed techniques should be chosen for a given application.
\item We do extensive simulations to show the benefit of using our framework compared to using a global fixed parameter, as well as highlight situations where one technique outperforms another. 
\item We describe the infrastructure required to put such a system in production for a large-scale social network platform. 
\item We show results on a real-world application that has resulted in significant metric wins.
\end{itemize}

The rest of the paper is organized as follows. In Section \ref{sec:problem}, we describe the generic problem formulation. We then describe our overall methodology in Section \ref{sec:method}. We illustrate our simulation studies in Section \ref{sec:simulation}, and discuss the system architecture for the proposed methods in Section \ref{sec:architecture}. We then deep dive into how our proposed methods were used in determining Notification send parameters at LinkedIn in Section \ref{sec:notificationApplication}, followed by some discussion around related work in Section \ref{sec:relatedwork} and concluding remarks and future directions in Section \ref{sec:discussion}.

\section{Problem Setup}
\label{sec:problem}

To describe the generic problem formulation, let us begin with some notational preliminaries. Let $J$ denote the total number of treatment variants under consideration. We assume that there are $K+1$ different metrics that we wish to track, where we have one primary metric and $K$ guardrail metrics. Let $C_1, \ldots, C_n$ denote the set of distinct, non-intersecting cohorts that exhibit heterogeneous behavior with respect to $K+1$ metrics under the treatment variants. With the member-level solution, a single member would represent a cohort $C_i$, and $n$ would be the total number of members. 

On each cohort $C_i$ for $i = 1, \ldots, n$, let $U_{i,j}^k$ denote the causal effect in metric $k$ by applying treatment variant $j$ vs control. We further assume that each such random variable $U_{i,j}^k$ is distributed as a Gaussian random variable having mean $\mu_{i,j}^k$ and standard deviation $\sigma_{i,j}^k$. For notational simplicity, for each metric $k = 0, \ldots, K$, let us denote $\Ub_k$ as the vectorized version of $U_{i,j}^k$ for $i = 1, \ldots, n$, $j = 1, \ldots, J$. Using a similar notation we get,
\begin{align*}
\Ub_k \sim N_{n,J} \left( \boldsymbol{\mu}_k, Diag( \boldsymbol{\sigma}_k^2)\right)
\end{align*}
Finally, let us denote $x_{ij}$ as the probability of assigning the $j$-th treatment variant to the $i$-th cohort and $\xb$ be its vectorized version. For ease of readability, we tabulate these in Table \ref{tab:notation}.

\begin{table}[th!]
\centering 
\begin{tabular}{|r | p{6cm} |}
\hline
 Symbol & Meaning \\ 
\hline
$J$ & Total number of treatment variants or choices.\\
\hline
$K$ & Total number of guardrail metrics\\
\hline
$C_i$  & $i$-th cohort (the smallest cohort would be a individual member) for $i = 1, \ldots, n$.\\
\hline
$\Ub_k$ & Vectorized version of $U_{i,j}^k$, which is the causal effect in metric $k$ by variant $j$ in cohort $C_i$. \\  
\hline
$\boldsymbol{\mu}_k$ & Mean of $\Ub_k$ \\  
\hline
$\boldsymbol{\sigma}_k^2 \Ib$ & Variance of $\Ub_k$\\  
\hline
$\xb$ & The assignment vector.\\
\hline
\end{tabular}
\caption{Notations used in the paper to formally describe various problem instances and solutions.}
\label{tab:notation}
\end{table}

Based on these notations, we can formulate our optimization problem. Let $k = 0$ denote the main metric. We wish to optimize the main metric keeping the constraint metrics at a threshold. Formally, we wish to get the optimal $\xb^*$ by solving the following:
\begin{equation}
\label{eq:example_main}
\begin{aligned}
& \underset{\xb}{\text{Maximize}} & & \xb^T\Ub_0\\
& \text{subject to}
& &\xb^T\Ub_k \leq c_k \qquad \text{ for } k = 1, \ldots, K. \\
&&&\sum_{j} x_{i,j} = 1\;\; \forall i,\qquad 0 \leq \xb \leq 1 \\
\end{aligned}
\end{equation}
where $c_k$ are known thresholds. Based on these, the overall problem can be formalized into two steps. 
\begin{enumerate}
\item[(i)] Identifying member cohorts $C_1, \ldots, C_n$ using data from randomized experiments, and then estimate the cohort-level causal effects $\Ub_k$. At a member-level set-up, where each member represents a cohort, we directly estimate the individual level causal effects.
\item[(ii)] Identifying the optimal selection of treatment variants $\xb^*$ by solving the optimization problem \eqref{eq:example_main}. 
\end{enumerate}

\section{Methodology}
\label{sec:method}
In this section, we describe our framework in detail. We first begin with how we can estimate heterogeneous causal effects at either cohort or member level. We then describe how we solve the optimization problem to select optimal treatment variants for each member. Since several layers of approximation exist in the two-step techniques, we discuss how to estimate the bias and variance of our optimal selection. In addition, we discuss the overall algorithm combining the two stages.

\subsection{Heterogeneous Effect Estimations}
\label{sec:causal}
Estimation of heterogeneous treatment effects is a well-studied topic in social and medical sciences \cite{Athey7353, stefan2015forest}. In this paper, we follow the potential outcomes framework from Rubin (1974) \cite{rubin1974estimating} and consider the following assumptions:
\begin{itemize}
\item \textit{Stable Under the Treatment Value Assumption} (SUTVA) \cite{rubin1974estimating}, which states that the response of the treatment unit only depends on the allocated treatment to that unit and not on the treatment given to other units. 
\item \textit{Stongly
Ignorable Treatment Assignment} \cite{rosenbaum1983central}, which combines the assumption of \textit{unconfoundedness} and \textit{overlap}. We refer to \cite{rosenbaum1983central} for the details.
\end{itemize}
Throughout this section, we assume that we have access to data from randomized experiments. That is, we have observed the values for each metric $k = 0, \ldots, K$, by giving treatment value $j$ to a randomized group of experimental units, for $j = 1, \ldots, J$. 

\subsubsection{Cohort-Level Heterogeneity}
To obtain cohort-level estimations, we use the recursive partitioning technique from \citet{Athey7353} to identify the heterogeneous cohorts. The Causal Tree Algorithm \cite{Athey7353} generates a tree and hence a partition of the entire set of members into disjoint cohorts $\{C_1^{j,k}, \ldots, C_{n}^{j,k}\}$. The superscript ${j,k}$ in the cohort set refers to treatment $j$ and metric $k$. 

To come up with a combined cohort definition by incorporating multiple treatments and metrics is a challenging problem. One option could be merging the  $J(K+1)$ tree models into one single cohort assignment. Various techniques can be used for this task \cite{Strecht2015}. However, most approaches would fragment the cohorts into very small subsets, and hence, the estimated treatment effects based on data in these sub-cohorts can be extremely noisy. We avoid this unwanted noise in the treatment effect estimation by \textit{carefully exploiting the within cohort homogeneity of the treatment effect}. We present this approach in Algorithm \ref{algo:mergeTree}, where we replace the double index $j,k$ by a single index $\ell$ for notational convenience. Since the within cohort homogeneity is one of the fundamental assumptions of the cohort level estimation approach, it is easy to see that the soundness of the treatment effect estimation in Algorithm \ref{algo:mergeTree} requires no additional assumption. 

In Algorithm \ref{algo:mergeTree}, we sequentially merge the cohort sets $\mathcal{S}_{j,k} = \{C_1^{j,k}, \ldots, C_{n}^{j,k}\}$ to obtain the following set of mutually exclusive and exhaustive cohorts
\[
\mathcal{S}_{out} = \big\{\cap_{j=1}^{J}\cap_{k=0}^{K} C^{j,k} \neq \emptyset \mid C^{j,k} \in \mathcal{S}_{j,k}\big\}.
\]

For each treatment $j$ and each metric $k$, we retain the estimated treatment effect and its variance from the original cohort $C^{j,k} \in \mathcal{S}_{j,k}$ in each sub-partition containing $C^{j,k}$. Since each $\mathcal{S}_{j,k}$ is exhaustive, this provides estimates of treatment effect and its variance for all sub-partitions, i.e.,
\[
\mathcal{T}_{out} = \{(\tau_{j,k}(C),~ \sigma_{j,k}^2(C)) \mid C \in  \mathcal{S}_{out},~  j = 1,\dots, J,~ k = 0, \ldots, K\}.
\]

\begin{algorithm}[!t]
\caption{Merging Trees}
\label{algo:mergeTree}
\begin{algorithmic}[1]
\Require  $L$ cohorts sets: $\{\{C_i^\ell\}_{i = 1}^{n_\ell} \mid \ell = 1,\dots, L\}$ and corresponding treatment effects and variances $\big\{ \{(\tau(C),~ \sigma^2(C)) \mid C \in \{C_i^\ell\}_{i = 1}^{n_\ell}\}  \mid \ell = 1,\dots, L\big\}$
\Ensure $\mathcal{S}_{out}$ and $\mathcal{T}_{out}$
\State Set $\mathcal{S}_{out} = \{C_i^1\}_{i = 1}^{n_1}$ and $\mathcal{T}_{out} = \{(\tau_{1}(C),~ \sigma_{1}^2(C)) \mid C \in  \mathcal{S}_{out}\}$
\For{ $\ell = 2, \ldots, L$ }
\For {$A \in \mathcal{S}_{out}$  }
\For {$B \in \{C_i^\ell \}_{i = 1}^{n_\ell}$}
\State $C = A\cap B$
\If {$C \neq \emptyset$}
\State $\mathcal{S}_{out} = \mathcal{S}_{out} \cup \{C\}$
\State $\mathcal{T}_{out} = \mathcal{T}_{out} \cup \{ (\tau_{m}(C),~ \sigma_{m}^2(C)) \mid m = 1,\ldots, \ell \}$, where 
\[
\tau_{m}(C),~ \sigma_{m}^2(C) = \left\{ 
\begin{array}{cc} 
\tau_{m}(A),~ \sigma_{m}^2(A) & \text{ for } m \leq l-1 \\
\tau_{m}(B),~ \sigma_{m}^2(B) & \text{ for } m=\ell 
\end{array}
\right.
\] 
\EndIf
\EndFor
\EndFor
\State $\mathcal{S}_{out} = \mathcal{S}_{out} \setminus \{A\}$
\State $\mathcal{T}_{out}  =  \mathcal{S}_{out} \setminus \{(\tau_{m}(A),~ \sigma_{m}^2(A)) \mid m = 1,\ldots, \ell - 1\}$
\EndFor
\end{algorithmic}
\end{algorithm}

\subsubsection{Member-level Heterogeneity}
To estimate the heterogeneous causal effects at a member level, we can choose any of the following options:
\begin{enumerate}
\item[(a)] \textbf{Causal Forest}: The Causal Forest Algorithm \cite{stefan2015forest} is an extension of the Causal Tree which was inspired by Random Forest Algorithm \cite{randomF} and use ensemble learning to incorporate results from multiple tree models. Here we consider all the $J(K+1)$ causal forests, and for each treatment $j$ and metric $k$, we get the member level effect from that specific forest model.
\item[(b)] \textbf{Two-Model Approach:} This is a baseline method (commonly applied in uplift modeling domain) that models the causal effect at a member level through the difference of the predicted response in the treatment and control models \cite{uplift}. Although this approach allows flexible choices of predictive models, the final member-level estimate $\hat{\tau}_{i} = \hat{Y}_{1,i}- \hat{Y}_{0,i}$, can be highly biased ($\hat{Y}_{1,i}$ and $\hat{Y}_{0,i}$ are the estimated potential outcomes).
\end{enumerate}

\subsection{Optimization Solution}
\label{sec:opt}
The optimization problem \eqref{eq:example_main} is stochastic since both the objective function and the constraints used in the optimization formulation are not deterministic but are coming from a particular distribution (Gaussian in the above case). More formally, we can rewrite the optimization problem in 
\eqref{eq:example_main} as:
\begin{equation}
\label{eq:stoc-opt}
\begin{aligned}
& \underset{\xb}{\text{Maximize}} & & f(\xb) = \Ebb( \xb^T \Ub_0 ) \\
& \text{subject to} & &  g_k(\xb) := \Ebb( \xb^T \Ub_k  - c_k) \leq 0, \qquad \; k = 1, \ldots, K. \\
& & &  \sum\nolimits_{j} x_{ij} = 1 \;\; \forall \; i , \qquad 0 \leq \xb \leq 1
\end{aligned}
\end{equation}
We can solve this problem via two routes. First, we can frame this as a pure stochastic optimization problem where the objective function $f$ and the constraints $g_j$ are not directly observed but observed via the realizations of the random variables $\Ub_0, \ldots, \Ub_K$. Second, we can frame it a deterministic problem where we replace the expectation with its empirical average. We discuss both in details below.

\subsubsection{Stochastic Optimization}
A method of solving a problem \eqref{eq:stoc-opt} is known as stochastic approximation (SA) \cite{robbins1951stochastic}, which has a projection step onto $\{g_k(\xb) \leq 0\}$ for $k = 1,\ldots, K$. This may not be possible in our situation. To have a more formal solution we generalize the Coordinated Stochastic Approximation (CSA) algorithm from \cite{lan2016algorithms}. The CSA algorithm in \cite{lan2016algorithms} solves the problem with a single constraint which we modify to work with multiple expectation constraints. Specifically, we start by framing the above problem as
\begin{equation}
\label{eq:main_problem}
\begin{aligned}
& \underset{\xb}{\text{Minimize}} & & f(\xb) := \Ebb_{ \Ub_0} \left(F(\xb, \Ub_0)\right)\\
& \text{subject to}
& & g_k(\xb) :=  \Ebb_{ \Ub_k} \left(G_j(\xb, \Ub_k)\right) \leq 0\qquad \text{ for } k = 1, \ldots, K. \\
&&& x \in \Xcal
\end{aligned}
\end{equation}
here $\Xcal$ denotes the set of non-stochastic constraints. 

Our algorithm, which we call Multiple Coordinated Stochastic Approximation (MCSA), is an iterative algorithm which runs for $N$ steps. At each step $t$ we start by estimating the constraint function. Specifically,  we simulate $\Ub_{k,\ell}$ for $\ell = 1, \ldots, L$ and estimate,
\begin{align}
\label{eq:Gfunc1}
\hat{G}_{k,t} = \frac{1}{L} \sum\nolimits_{\ell = 1}^L G_k(\xb_t, \Ub_{k,\ell}).
\end{align}
Then if all the estimated constraints $\hat{G}_{k,t}$ are less than a threshold $\eta_{k,t}$, we choose our gradient to be the stochastic gradient of the objective function, $F'(\xb_t, \Ub_{0,t})$. Otherwise, from the set of violated constraints $\{k : \hat{G}_{k,t} > \eta_{k,t}\}$, we choose a constraint $k^*$ at random and consider the gradient to be the stochastic gradient of that constraint, $G_{k^*}'(\xb_t, \Ub_{k^*,t})$. We then move along the chosen gradient for a particular step-size $\gamma_t$ and apply the traditional proximal projection operator to get the next point $\xb_{t+1}$. Our final optimal point is the weighted average of those $\xb_t$ where we have actually travelled along the gradient of the objective. The overall algorithm to solve \eqref{eq:main_problem} is now written out as Algorithm \ref{algo:mcsa}.

\begin{algorithm}[ht]
\caption{Multiple Cooperative Stochastic Approximation}\label{algo:mcsa}
\begin{algorithmic}[1]
\State Input : Initial $\xb_1 \in \Xcal$, Tolerances $\{\eta_k\}_t, \{\gamma\}_t$, Iterations $N$
\For{ $t = 1, \ldots, N$ }
\State Estimate $\hat{G}_{k,t}$ for all $k \in 1,\ldots, K$ using \eqref{eq:Gfunc1}.
\If{$\hat{G}_{j,t} \leq \eta_{j,t} \text{ for all } j$}
\State Set $h_t = F'(\xb_t, \Ub_{0,t})$
\Else
\State Randomly select $k^*$ from $\{k : \hat{G}_{k,t} > \eta_{k,t}\}$
\State Set $h_t = G_{k^*}'(\xb_t, \Ub_{k^*,t})$
\EndIf
\State Compute $\xb_{t+1} = P_{\xb_t}(\gamma_t h_t)$
\EndFor
\State Define $\Bcal = \left\{ 1 \leq t \leq N  : \hat{G}_{k,t} \leq \eta_{k,t} \; \forall k \in \{1, \ldots, K\} \right\}$
\State \Return $\hat\xb := \frac{\sum_{t \in \Bcal} \xb_t \gamma_t}{\sum_{t \in \Bcal} \gamma_t}$
\end{algorithmic}
\end{algorithm}
 
It has been theoretically proven in \cite{lan2016algorithms} that if there was a single stochastic constraint, then the number of iterations $N$ required to to come $\epsilon$ close to the optimal both with respect to the objective and the constraint is of the order $\Ocal(1/ \epsilon ^2)$. Following a very similar proof strategy it is not hard to see that we can carefully choose the tolerances $\{\eta_k\}_t, \{\gamma\}_t$ such that if we follow Algorithm \ref{algo:mcsa} for $N$ steps, then
\begin{align*}
\Ebb(f(\hat{\xb}) - f(\xb^*)) \leq \frac{c_0}{\sqrt{N}} \;\;\text{ and } \;\; \Ebb(g_k(\hat{\xb}))  \leq \frac{c_k}{\sqrt{N}}  \;\; \forall k \in \{1,\ldots, K\}
\end{align*}
We do not go into the details of the proof in this paper for brevity. Please see \cite{lan2016algorithms} for further details.

\subsubsection{Deterministic Optimization}
The method of solving \eqref{eq:stoc-opt} in a deterministic sense is known in literature as \textit{sample average approximation} (SAA) \cite{kleywegt2002sample}. In SAA, instead of solving the stochastic problem, we replace the stochastic objective and the stochastic constraints via their empirical sample expectation.  Namely, we replace the mean treatment effects $\Ebb(U_k)$ with estimated treatment effects $\hat{\boldsymbol{\mu}}_k$. That is, the problem becomes
\begin{equation}
\label{eq:det-opt}
\begin{aligned}
& \underset{\xb}{\text{Maximize}} & & f(\xb) = \xb^T \hat{\boldsymbol{\mu}}_0 \ \\
& \text{subject to} & &  g_k(\xb) :=  \xb^T  \hat{\boldsymbol{\mu}}_k  - c_k \leq 0, \qquad \; k = 1, \ldots, K. \\
& & &  \sum\nolimits_{j} x_{ij} = 1 \;\; \forall \; i , \qquad 0 \leq \xb \leq 1.
\end{aligned}
\end{equation}
This problem can be solved by any off-the-shelf Linear Programming (LP) solver. However, observe that this problem makes no use of the variance of the random variables. In many cases, such as ours, the SAA solution might lead to an infeasible problem, due to the large measurement error, especially in cases with large variance. Moreover, SAA is computationally challenging for a general function $g_j$. 



\subsection{Bias and Variance of Optimal Assignments}
In practice, the means $\mu^k_{i,j}$ and variances $\left(\sigma^k_{i,j}\right)^2$ are unknown.  When performing the stochastic (or even deterministic) optimization routine, these quantities are replaced by sample estimates $\hat \mu^k_{i,j}$ and variances $\left(\hat \sigma^k_{i,j}\right)^2$ which introduces error in the estimation of the optimal assignment ${\bf x}^*$.  The variability and bias of the estimate, say $\hat x$ of the optimal assignment can be estimated using bootstrap as follows. 

Assume that we have $N$ users.  Randomly resample $N$ members from the original set of members, with replacement.  Recompute the assignment by solving problem \eqref{eq:main_problem} or \eqref{eq:det-opt} using the means and variances estimated on the selected members.  The variance of these bootstrap assignments provides an estimate of variance.  Similarly, the difference between the original assignment and the mean of the bootstrap assignments estimates the bias.  

Of course, in many applications, resampling members is an unnecessary computational burden.  
Assume that in cohort $C_i$ there are $N_{i,j,treat}$ members given treatment $j$ and $N_{i,cont}$ members given the control and define $N_{i,j} = N_{i,j,treat} + N_{i,cont}$.  Further assume that in cohort $C_i$, the variance in the metric $k$ by applying treatment $j$ is $\left(\hat \sigma^k_{i,j, treat} \right)^2$ and the variance for the control is $\left(\hat \sigma^k_{i,cont}\right)^2$ so that the variance $\left(\hat \sigma^k_{i,j}\right)^2$ can be written as $$\left(\hat \sigma^k_{i,j}\right)^2 = \left(\hat \sigma^k_{i,j, treat}\right)^2 + \left(\hat \sigma^k_{i,cont}\right)^2.$$
The mean computed on the resampled members will be distributed as $N(\hat \mu^k_{i,j}, \left(\hat \sigma^k_{i,j}\right)^2/N_{i,j})$, and the resampled variance will be distributed as $\left( \hat \sigma^k_{i,j, treat}\right)^2 \cdot \Xcal  + \left(\hat \sigma^k_{i,cont}\right)^2\cdot \Xcal'$ where $\Xcal$ and $\Xcal'$ are independent chi-squared random variables with one degree of freedom.  Rather than resampling members, bootstrap estimates of the optimal assignment can be attained by computing the assignment across many simulations of these means and variances, as summarized in Algorithm \ref{algo:bootstrap}.  Note that this algorithm is stated for the stochastic optimization problem.  An obvious, and simpler analog holds for the deterministic optimization case, where only the means need to be resampled.  
\begin{algorithm}[ht]
\caption{Efficient Bootstrap Bias and Variance Estimation}\label{algo:bootstrap}
\begin{algorithmic}[1]
\State Input : Initial assignment $\hat x$, Bootstrap iterations $B$
\For{ $b = 1, \ldots, B$ }
\State Sample $Z_{i,j,k} \sim N(0,1)$, $X_{i,j,k} \sim \chi^2_{(N_{i,j,treat}-1)}$, and $X_{i,k} \sim  \chi^2_{(N_{i,cont}-1)}$ independently
\State Compute $$\hat \mu^k_{i,j,b} = \hat \mu^k_{i,j} + \frac{\hat \sigma^k_{i,j}}{N_{i,j}} Z_{i,j,k}$$ and $$\left(\hat \sigma^k_{i,j,b}\right)^2 = \left(\hat \sigma^k_{i,j, treat}\right)^2 \cdot \frac{X_{i,j,k}}{N_{i,j,treat}}  + \left(\hat \sigma^k_{i,cont}\right)^2\cdot \frac{X_{i,k}}{N_{i,cont}}$$ 
\State Calculate the $b$-th bootstrap assignment $\hat {\bf x}_b$ by solving Equation \eqref{eq:main_problem} with the resampled means and variances.  
\EndFor
\State Compute the mean of the bootstrap estimates, $\bar {\bf x} = \sum \hat {\bf x}_b$
\State Find bootstrap variance 
$$\widehat{var}(\hat x) = \frac{1}{B}  \sum (\hat {\bf x}_b - \bar {\bf x} )(\hat {\bf x}_b - \bar {\bf x} )'$$ and the bootstrap bias $$\frac{1}{B} \widehat{bias}(\hat x) = \frac{1}{B}\sum \hat {\bf x}_b - \hat x.$$
\State \Return $\widehat{var} (\hat x)$ and $ \widehat{bias} (\hat x)$
\end{algorithmic}
\end{algorithm}

The bootstrap estimates of bias can be used to bias correct the original parameter estimates by subtracting the estimated bias from the original assignment. There are two advantages of the bias-corrected assignments.  The first is that they provide a better estimate of the solution to the optimization problem \eqref{eq:main_problem} in the sense of expected treatment effect.  To see this, assume that $\hat x$ is a biased estimator and $\hat x_u$ is an unbiased estimator, then 
\begin{align*}
\Ebb(\hat x' U) &= \Ebb\left(\Ebb(\hat x' U|U)\right) =\Ebb(\Ebb(\hat x')U) \\
& \leq  \Ebb(({\bf x}^*)'U) =  \Ebb(\Ebb(\hat x_u' U|U)) =\Ebb(\hat x_u'U)
\end{align*}
The second is that in examples where the treatment effects have large variance (especially when the estimated treatment effects are not significantly different than zero), the original assignments can overweight treatments which had large effects due to chance rather than an inherent improvement from the treatment.  The bias-corrected estimates down-weight such instances, and overall provide a more conservative assignment. 

\begin{remark}
Bias correction may result in an assignment that is not in the probability form but this is easily remedied by bias correcting on the log-odds scale.
\end{remark}

\subsection{Overall Algorithm}
\label{sec:overall}
Our end-to-end framework for estimating cohort or member level causal effects and using these to find optimal treatment assignments is given in Algorithm \ref{algo:overall}.

\begin{algorithm}[ht]
\caption{: Optimal Treatment Selection}\label{algo:overall}
\begin{algorithmic}[1]
\State Run Randomized Experiment to collect data across various treatment variants and metrics.
\State Generate a cohort-level or member-level causal effect for the different parameters using the technique in Section \ref{sec:causal}. 
\State Solve the optimization problem (stochastic or deterministic) as given in \ref{sec:opt}.
\State Return bias corrected assignment $\hat{x}$ by following Algorithm \ref{algo:bootstrap}.
\end{algorithmic}
\end{algorithm}

Note that steps 2 and 3 of Algorithm \ref{algo:overall} each have two options, yielding four distinct approaches as shown in Table \ref{tab:compare}. Although the deterministic optimization has obvious computational advantages over the stochastic counterpart, it may not be the right choice depending on what was done in step 2. Especially when using the cohort-level effect estimation, stochastic optimization is generally preferable because it provides assignments that are reflective of the inherent uncertainty in the causal effect estimates. 

In the context of the member-level approaches, computation of the variances would generally rely on the correct specification of the models, and consequently, the variance estimates are not usually reliable. When this is the case, the issue of incorrect variance estimation can be avoided by simply using the deterministic optimization. That being said, if we are considering the computation complexity of the problem, a cohort-based solution is always preferable, since solving a member-level optimization problem for a large-scale social network having hundreds of millions of members can be extremely challenging. Our recommendations regarding choice of methodologies is summarized in Table \ref{tab:compare}. Next, we use a simulation setup to study the performance and start quantifying the benefits of the proposed algorithms under different conditions.

\begin{table}[ht]
\centering
\begin{tabular}{||c|c|c||}
\hline
Heterogeneous & \multicolumn{2}{c||}{Optimization}\\
Causal Effect & Stochastic & Deterministic\\
\hline
\multirow{2}{*}{Cohort-Level} & Preferred & Not-Preferred \\
& Scalable & Scalable \\
\hline
\multirow{2}{*}{Member-Level} & Not-Preferred & Preferred \\
& Non-Scalable & Hard to Scale\\
\hline
\end{tabular}
\caption{Comparison of different methods for optimal treatment selection}
\label{tab:compare}
\end{table}
\section{Simulation Analysis}
\label{sec:simulation}
We conduct a simulation analysis to examine the proposed methodologies (enumerated below) in different scenarios and compare their performance with a global allocation baseline (denoted as $Global$).

\begin{enumerate}
\item  A heuristic cohort-level solution (cohorts generated by binning the heterogeneity variable features based on their quantile values) paired with stochastic optimization which served as a baseline to compare with model-based cohort identifications (denoted as $HT.ST$).
\item cohort-level estimations using Causal Tree model paired with stochastic optimization (denoted as $CT.ST$). To incorporate heterogeneity from all treatments and metrics, we merged the causal trees using Algorithm 1.
\item member-level estimations with Causal Forest model  \cite{stefan2015forest} paired with deterministic optimization (denoted as $CF.DT$).
\item member-level estimations with Two-Model approach (applied Random Forest models \cite{randomF} for outcome predictions) paired with deterministic optimization (denoted as $TM.DT$).
\end{enumerate}
 
\subsection{Simulation Data Generation}
\begin{figure}[ht]
\centering
  \includegraphics[width=0.65\linewidth]{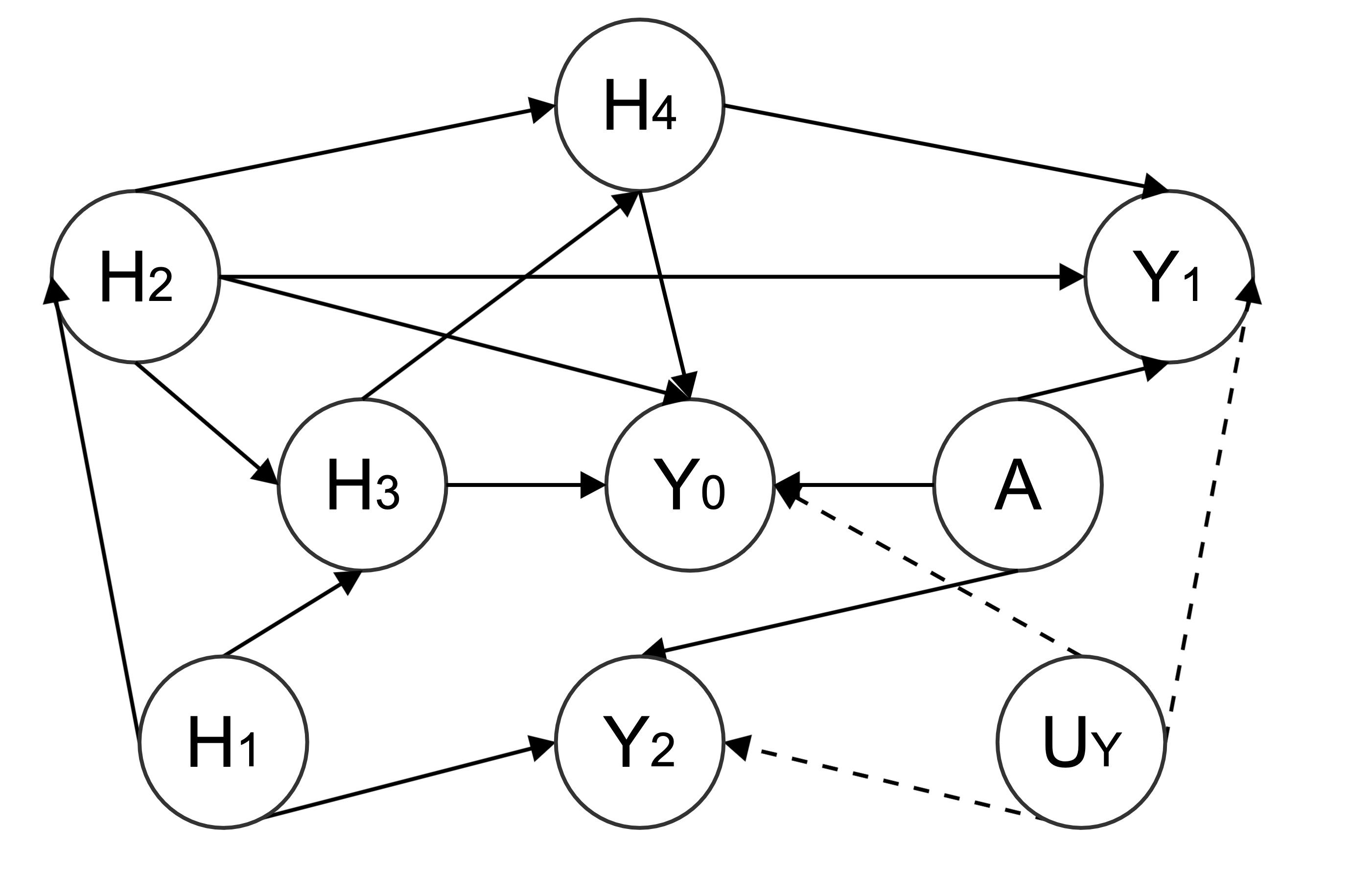}
  \caption{Simulation causal graph with treatment variables (A), heterogeneity factors (H), latent variables (U) and metrics of interest (Y). The edges encapsulate the causal functions including the amount of uncertainty.}
  \label{fig:simulationCausalGraph}
\end{figure}

We leverage \texttt{simcausal} R package \cite{simcausal} to generate simulation datasets for both training and testing purposes under self-defined causal Directed Acyclic Graphs (DAG). In a DAG, causal relationships are represented by directed arrows between the nodes, pointing from cause to effect \cite{simcausal}. We still assume there exists $k = 0, \ldots, K$ metrics, and $j = 1, \ldots, J$ treatment values. We define $A_j$ as the treatment variables, $Y_k$ are the metrics,  $U_Y$ as a latent variable impacts $Y_k$, and $H_m$ as the heterogeneous variables. We simulate heterogeneity by introducing interaction terms between $A_j$ and $H_m$ on $Y_k$. We assume the treatments all satisfy the same causal DAG set up (e.g., see Figure \ref{fig:simulationCausalGraph}), but the weights on the causal relationship are slightly different. Each $A_j$  is associate with a different set of weight vectors $W_{j,k}$ and the metric $Y_{k} = \sum_{j = 1}^{J}({W_{j,k}(A_j(H_{m,j})^2 + U_Y)})$. 

In order to measure the performance of the proposed methods with different levels of noise in data, we introduce an uncertainty weight hyperparameter which is a multiplicative factor on the standard deviation of the associated variables ($U_Y$ and $Y_k$). 

\begin{figure*}[th]
    \centering
    \begin{subfigure}[t]{\textwidth}
    \centering
         \begin{subfigure}[t]{0.33\textwidth}
                 \renewcommand\thesubfigure{\alph{subfigure}}
        		\includegraphics[width=\linewidth]{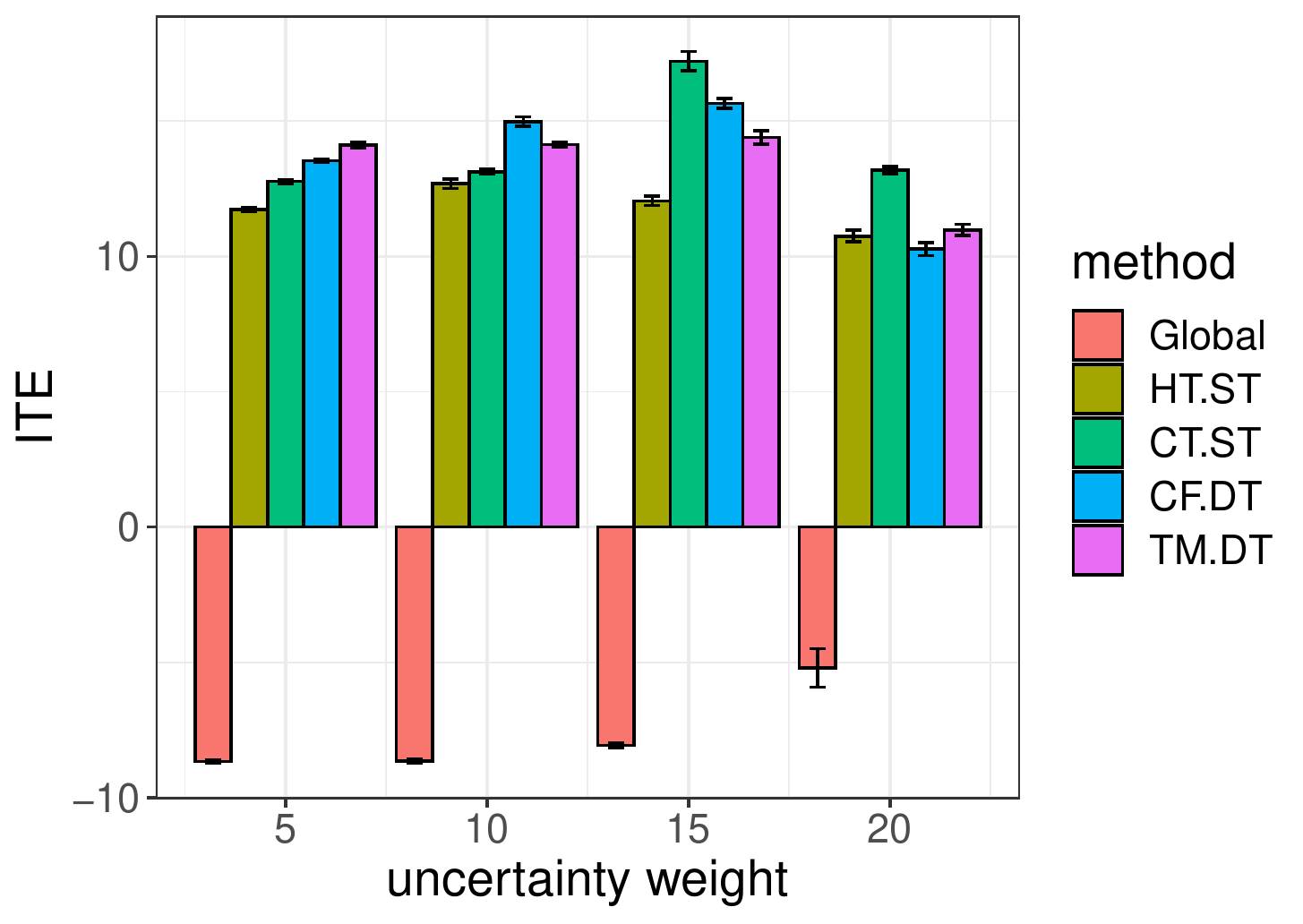}
  		\caption{Evaluation on the objective metric $Y_0$}
  		\label{fig:y0}
    	\end{subfigure}%
	~ 
        \begin{subfigure}[t]{0.33\textwidth}
                 \renewcommand\thesubfigure{\alph{subfigure}}
        		\includegraphics[width=\linewidth]{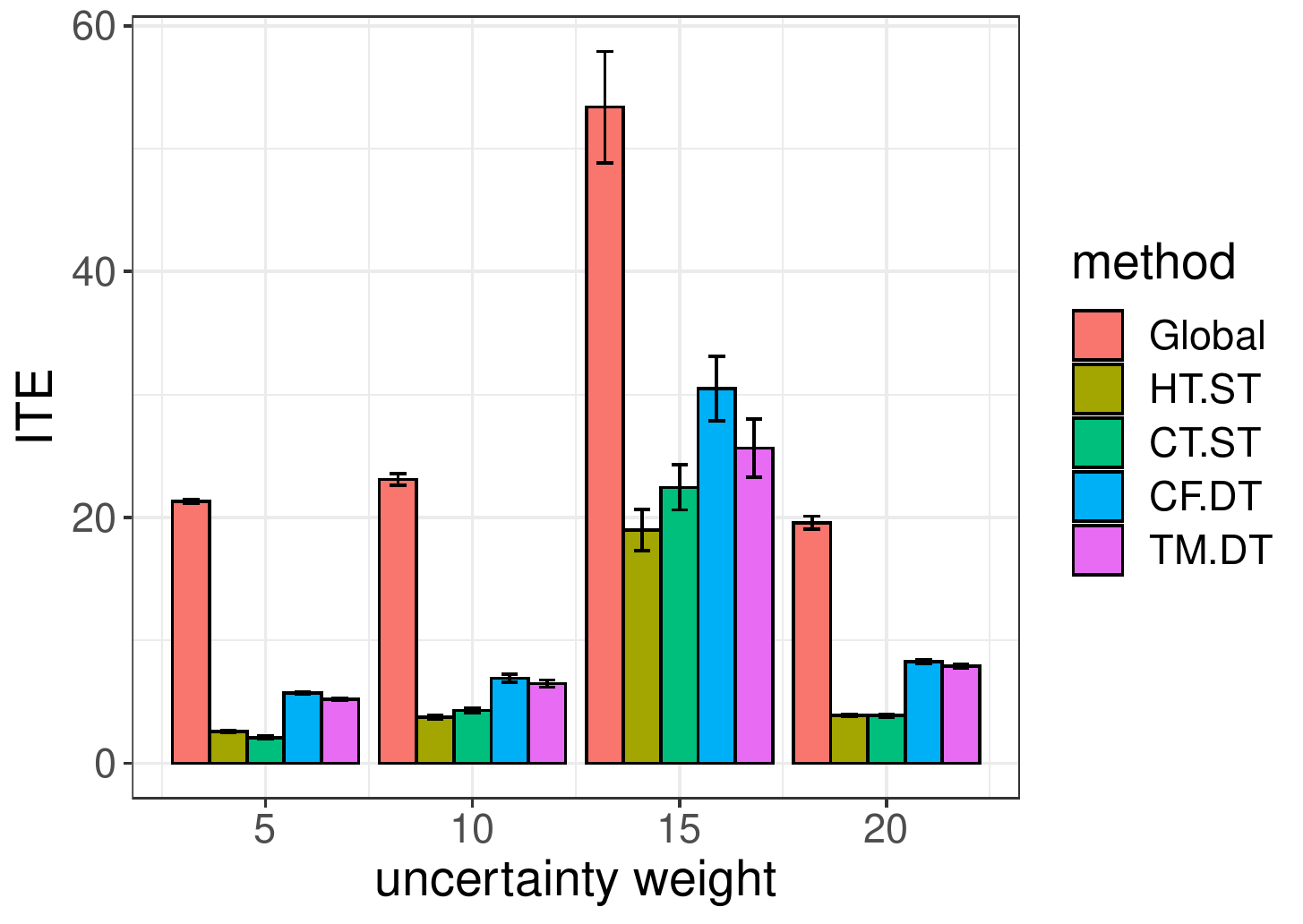}
  		\caption{Evaluation on the constraint metric $Y_1$}
  		\label{fig:y1}
    	\end{subfigure}%
	~ 
    	\begin{subfigure}[t]{0.33\textwidth}
        	\centering
	        \renewcommand\thesubfigure{\alph{subfigure}}
       		\includegraphics[width=\linewidth]{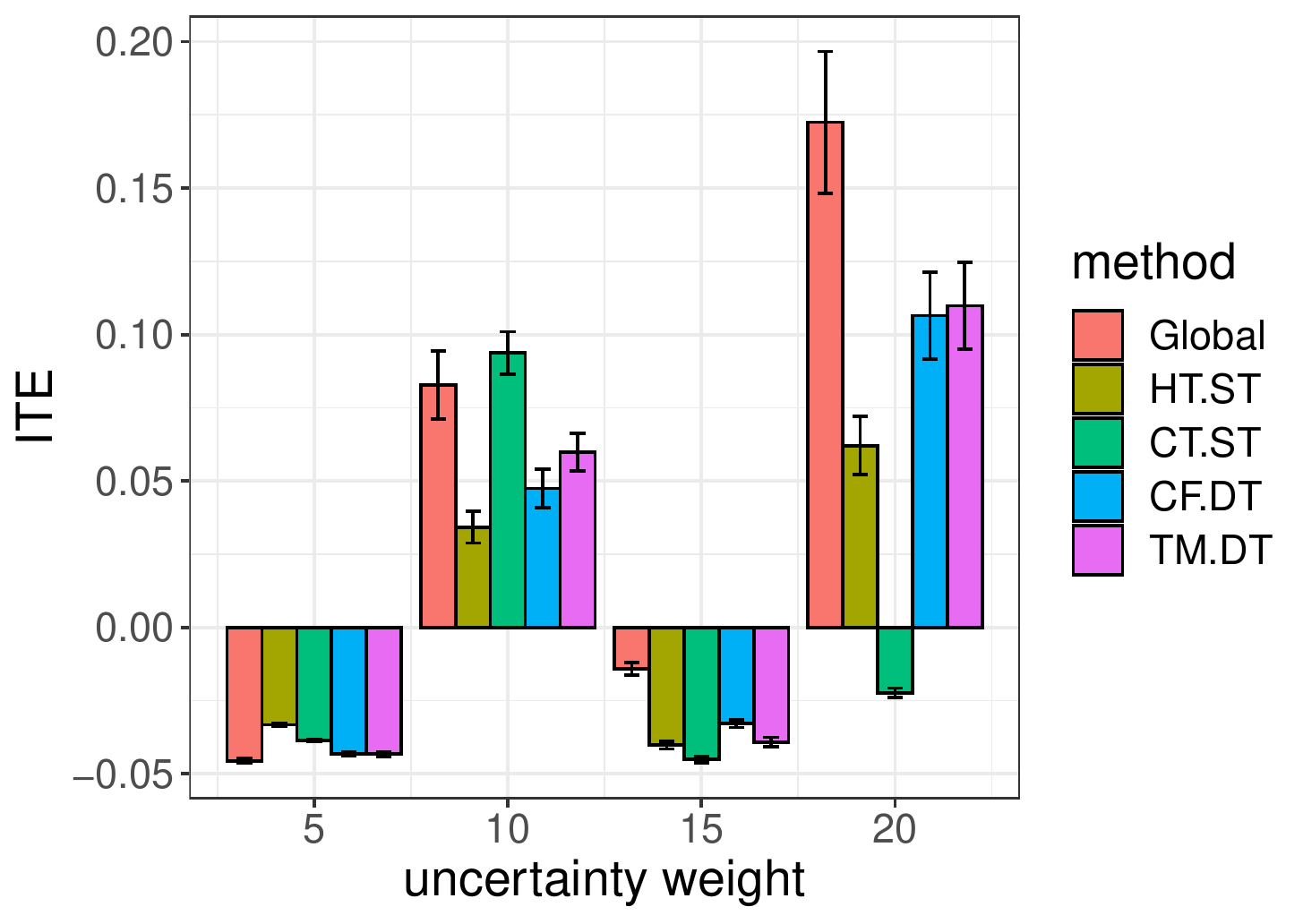}
  		\caption{Evaluation on the constraint metric $Y_2$}
  		\label{fig:y2}
   	\end{subfigure}
   \end{subfigure} 
    \caption{ Simulation results on comparing the proposed methods with different noise levels (uncertainty weights), where x-axis represents the uncertainty weight, and y-axis represents the mean of individualized treatment effects (ITE) normalized by the average outcome of control group (error bar show the standard deviation across different runs) on the simulated test datasets.} 
      \label{fig:evaluationResults}

\end{figure*}

\subsection{Simulation Analysis and Results}
We then evaluate the policy $\xb^*$ generated based on each proposed methods ($CT.ST$, $CF.DT$, $TM.DT$) compared with a heuristic cohort-level method ($HT.ST$) and the best solution from one of the global parameter setting ($Global$) using simulated datasets. They are generated based on a DAG as Figure \ref{fig:simulationCausalGraph} where there are $J = 3$ treatment values and one objective metric $Y_0$ and two constraint metrics $Y_1$ and $Y_2$ (where number of guardrail metrics $K=2$). 

We set the optimizations to maximize the lifts on $Y_0$, and while holding the impacts on $Y_1$ and $Y_2$ close to 0 (for all three metrics, we desire them to increase along the positive direction). For the $HT.ST$ solution, we created 16 cohorts by binning the heterogeneous variables ($H_m$ where $m = 1,\ldots,4$) based on their median values. In order to make the proposed methods more comparable, we use Random Forest models \cite{randomF} trained with the same set of features in the Two-Model approach ($TM.DT$). 

For each noise level, we calculate the optimal solution $\xb^*$. Note that the optimal allocation can be probabilistic across the $J=3$ treatments, that is $\xb_i^* = (x_{i,1}, \ldots, x_{i,3})$ such that $\sum_j x_{i,j} = 1$. We then consider the normalized mean of individualized treatment effect (ITE) for each metric $k$ at optimal $\xb^*$ as
$$
\tau(\xb^*)_k =\frac{\frac{1}{n} \sum\nolimits_{i=1}^n \sum\nolimits_{j=1}^3 (Y_{1,i,k} - Y_{0,i,k}) x_{i,j}}{\mu_{0,k}}
$$
where, $Y_{1,i,k}$ is the outcome for metric $k$ if the member $i$ receive treatment $j$, $Y_{0,i,k}$ is the outcome if the member receive the control set-up. $(Y_{1,i,k} - Y_{0,i,k})$ represents the individualized treatment effect. We normalize the ITE by the control group mean $\mu_{0,k}$ to make results comparable across different simulated datasets. We repeat this process 10 times and report the average and standard deviation for each noise level in Figure \ref{fig:evaluationResults}.

To simulate based on a realistic scenario where objective metrics usually move in the opposite direction to some constraint metrics, we randomly generate the weights $W_{j,k}$. As shown in  Figure \ref{fig:evaluationResults}, the global parameter setting ($Global$) can not achieve a similar level of lifts on the objective metric $Y_0$ while holding all the constraints within bounds (changes in $Y_1$, $Y_2$ being close to zero). All the proposed approaches perform better than the $Global$ solution in balancing between the objective and the constraints.

The Causal Tree based cohort identification approach paired with stochastic optimization ($CT.ST$) significantly outperforms the heuristic solution ($HT.ST$). The method consistently achieved higher gains on the objective metric which demonstrates the value of smart cohort identifications through models.

Comparing between the cohort vs. member level methods, we observe that with low noise levels (uncertainty weight), member-level solutions ($CF.DT$ and $TM.DT$) perform better than the cohort-level solution ($HT.ST$, $CT.ST$). Member-level solutions allow more advanced algorithms (e.g. ensemble models) to generate more personalized estimations and should be the oracle solution with zero noise in the system. However, along with an increase in the noise level, $CT.ST$ quickly starts to catch up and can outperform the member-level solutions with a high noise setting. This is because the stochastic optimization handles variance much better than deterministic optimization. We also replicate the simulation analysis under some different causal DAG settings, and observed similar patterns as shown in Figure \ref{fig:evaluationResults}. Now that we have a more quantitative understanding and validation of our methods, we provide some pointers on how to set up such a solution in practice.

\section{System Architecture}
\label{sec:architecture}
\begin{figure}[!ht]
  \centering
  \includegraphics[width = 1.15\linewidth]{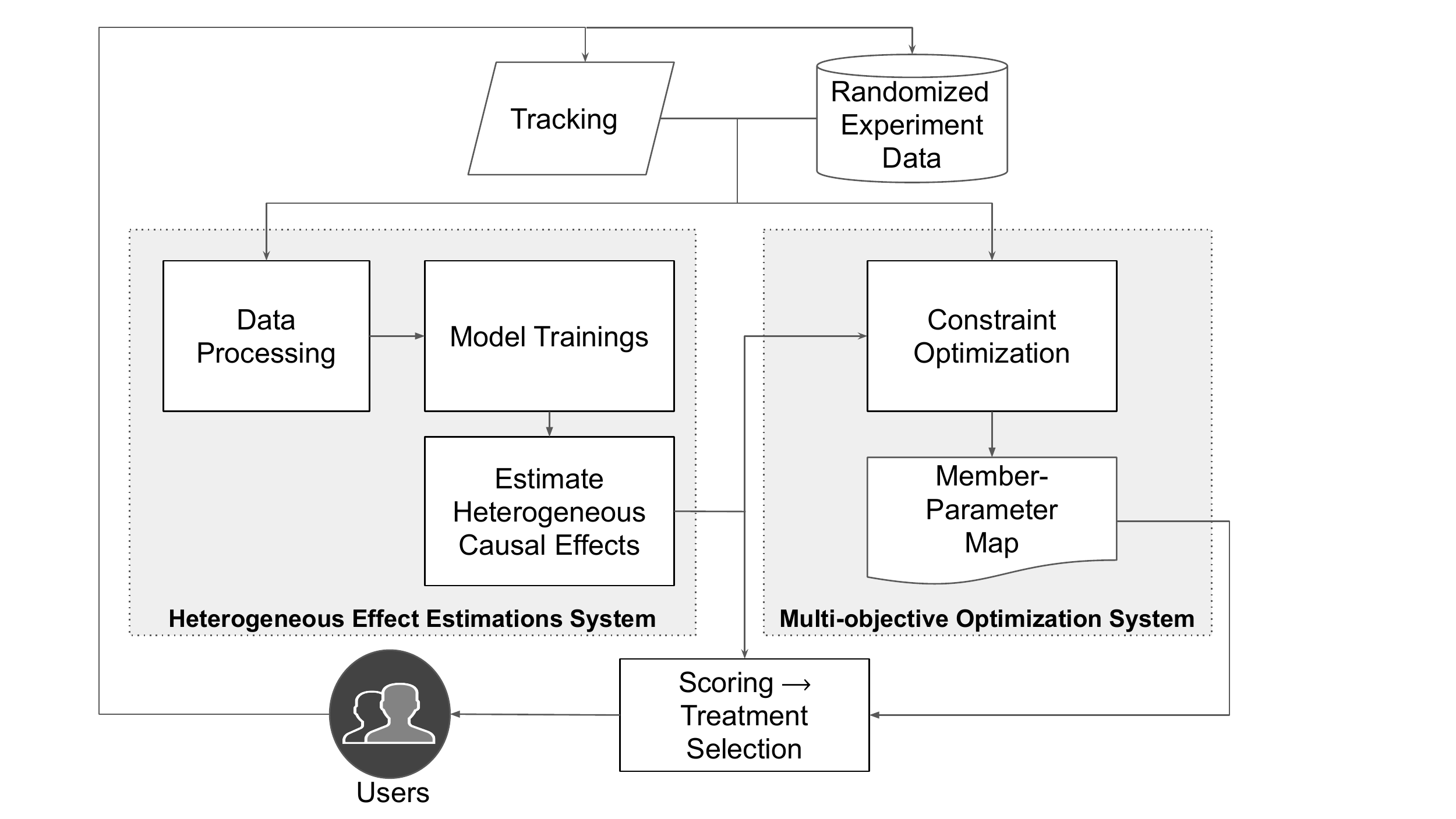}
  \caption{System architecture showing the two key modules of heterogeneous effect estimation and optimization, along with the scoring and tracking modules that complete the feedback loop.}
  \label{fig:systemArchitecture}
\end{figure}

We outline a general engineering architecture for the personalized treatment selection in Figure \ref{fig:systemArchitecture}. It consists of two major components: one for heterogeneous causal effect estimations and the other for the optimization module. Both components leverage tracking data (metrics and features) and randomized experiment data (i.e., the allocation of users to various treatment variants).  

The heterogeneous effect estimation system is similar to a standard response prediction training system where it is composed of the following pipelines:
\begin{itemize}
\item A flow for processing tracking and randomized test data to generate training and cross-validation sets,
\item A training pipeline to generate the estimations for each treatment variant $j$ and each metric of interest $k$. For the cohort-level solution, we also convert the tree model as a set of decision rules for clustering members into disjoint cohorts.
\end{itemize}

The output of the first system (either a finalized cohort definition or all associated models) together with the tracking and experiment allocation data serves as the inputs for the optimization system. In this stage, we estimate the cohort or member level effects for each variant $j$ and metric of interest $k$ and fit the estimations in the optimization pipeline to generate the final member-parameter mapping. 

We suggest pairing member-level estimations with deterministic optimization, and cohort-level estimations with stochastic optimization as illustrated in Section \ref{sec:overall}. For the cohort-level solution, the system produces the cohort definition and the mapping between cohorts and selected treatment variants which can be encoded in a model file. Applying the personalized treatment selection in an online system only requires a light-weight tree model scoring (since cohorts are identified using causal trees or forests) and post-processing flows to assign the selected variant. The solution would apply to a wide range of online systems without introducing a significant level of additional latency. For the member-level solutions, a large-scale Linear Programming (LP) optimization solver is required to solve for the optimal policy $\xb^*$ and hence should be avoided for extreme-scale applications due to the computational complexity. Next, we describe our experience using this proposed architecture to improve notification sending decisions on the LinkedIn platform.

\section{Notification System at LinkedIn}
\label{sec:notificationApplication}

\begin{figure}[!ht]
        \centering
        \includegraphics[height = 2.8in]{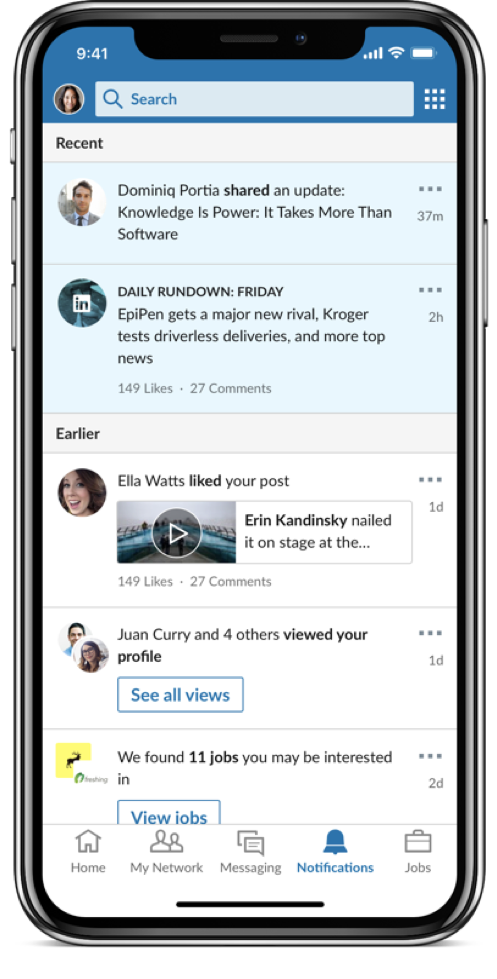}
        \caption{The notifications tab in the LinkedIn mobile app. Notifications have become a critical way to engage with users by proactively informing them about important events.}
        \label{fig:exnoti}
\end{figure}

Continuously growing mobile phone usage has caused notifications to become an increasingly critical channel for social networks seeking to keep members informed and engaged. Figure \ref{fig:exnoti} showcases a mobile notification tab page that LinkedIn uses to inform members about various activities (e.g., shares, news, birthdays) happening in their social network as well as relevant job recommendations \cite{yannotification2018}. This Notifications system is underpinned by a response prediction model that learns members' preferences based on their engagement patterns. 

To avoid overwhelming members, a hard limit (daily volume cap) was introduced in the system to limit the number of notifications a member can receive in a day. However, members can have different preferences for notification volume, and any global choice for such a cap can lead to sub-optimal member experience. Furthermore, as a member's network and its associated content ecosystem grow, there is a richer set of information pertinent to them. In the sections below, we present the application of our proposed method to learn personalized notification volume caps per member. Notification volume cap is an excellent example of a critical parameter in a recommender system, which is traditionally not learned and often set via ``global allocation''. Online A/B tests comparing personalized cap selection against a heuristic cohort-level solution and the global allocation baseline showed significant positive impact across multiple engagement metrics, indicating the effectiveness of our approach in providing value to hundreds of millions of LinkedIn members. Next provide some details on how we built the notifications solution to aid and motivate practitioners to adopt this in their systems.

\subsection{Personalized Capping Problem} 
Notifications are an important driver for member visits and engagement to LinkedIn. Sending more notifications can increase visits, but it also has negative consequences in terms of members engaging less with notifications (reduction in click-through rate) and increase in notifications disables. The system initially had a fixed cap which was the same for all LinkedIn members. This cap determined the maximum number of notifications a user could receive in a 24-hour window. Our goal with introducing personalized volume caps is to maximize visits to LinkedIn with constraints on click-through rate and Notification disables metrics (detailed descriptions in Table \ref{tab:notmetrics}). 

\subsection{Data Collection}
Data is gathered through a conventional A/B experiment, randomly assigning different volume caps to a small sample of users. We collect data for two weeks. In addition to the metrics mentioned in Table \ref{tab:notmetrics}, we include four major categories of member features in the heterogeneous cohorts learning process:

\begin{itemize}
\item Member profile features, such as country, preferred language, job title, etc.
\item Notification specific features for the member, such as past notifications and emails received. 
\item Other activity features including past visits, feed consumptions, likes, shares, comments, etc.
\item Features from members' networks (such as the number of connections/followers).
\end{itemize}

\begin{table}[h]
\centering
\begin{tabular}{p{2.5cm}|p{5.4cm}} 
\hline\hline 
Metrics & Descriptions
\\ [0.5ex]
\hline 
Sessions (Objective) & Number of visits to the LinkedIn site/app\\
\hline
Notification Sends & Volume of notifications sent to members\\
\hline
Notification CTR & Click through rate on notifications \\
\hline
Total Disables & Number of total disables on notifications  \\
\hline 
\end{tabular}
\caption{Metrics of Interest for Personalized Capping} 
\label{tab:notmetrics}
\end{table}

\subsection{Implementations}
We observed a high level of noise in members' responses to the various cap parameter changes in the data we gathered. Our offline simulations suggested that the cohort-level solution paired with stochastic optimization ($CT.ST$) performs best in this setting, so we decided to implement this strategy online. We developed the whole system offline using Apache Spark for data processing and scoring to generate final policy regularly, and leveraged packages in \texttt{R} for causal tree model training and stochastic optimization. 

\begin{remark}
We have explored the merging tree approach thoroughly in simulation studies. A simpler alternative that worked for us in practice was to use a causal tree identified using the objective metric since, in our application, the causal heterogeneity was much more pronounced on the objective metric.
\end{remark}

\subsection{Offline Analysis}
\label{sec:experiments}

\begin{figure}[ht]
    \centering
    \begin{subfigure}[t]{0.48\textwidth}
    \centering
        \begin{subfigure}[t]{0.5\textwidth}
                 \renewcommand\thesubfigure{\alph{subfigure}}
        		\includegraphics[width=\linewidth]{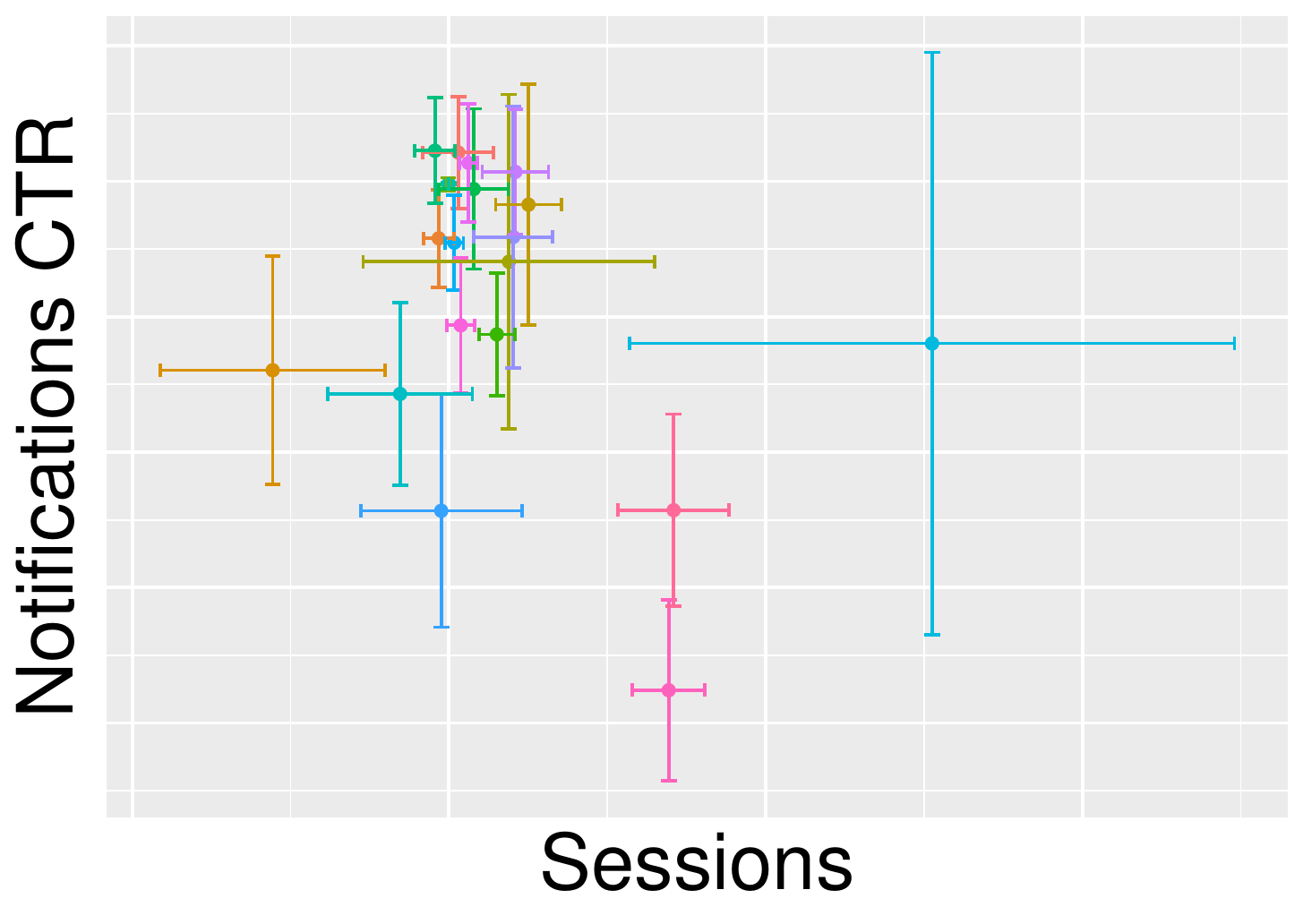}
  		\caption{Heterogeneous Cohorts}
  		\label{fig:pepperCohorts1}
    	\end{subfigure}%
	~ 
    	\begin{subfigure}[t]{0.5\textwidth}
        	\centering
	        \renewcommand\thesubfigure{\alph{subfigure}}
       		\includegraphics[width=\linewidth]{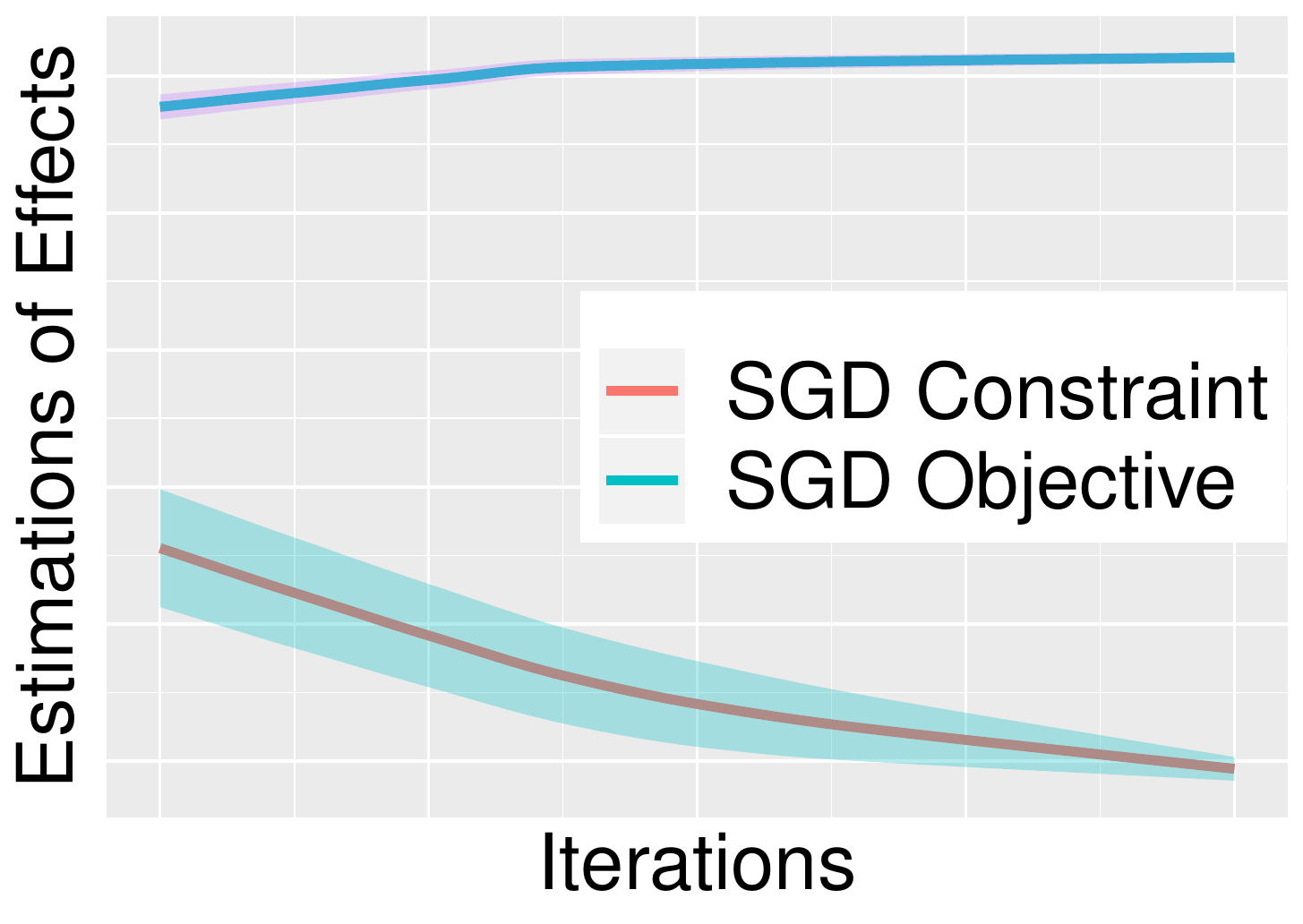}
  		\caption{Expectated Sessions Lift}
  		\label{fig:prophetsp}
   	\end{subfigure}
   \end{subfigure}

    \caption{ Heterogeneous cohorts in Notifications, where each color in (a) corresponds to one cohort. Figure (b) show the expected lift in sessions and expected decrease in disables as function along iterations of the Algorithm \ref{algo:mcsa}. } 
\end{figure}

Figure \ref{fig:pepperCohorts1} visualizes the cohorts generated using an optimized value of $\alpha$ which balances the objective in the Causal Tree algorithm \cite{Athey7353}. Our experiments show that a lower value of $\alpha$ leads to smaller number of cohorts and lower variance (hence higher confidence) on the cohort-level effect estimations; and vice-versa. We select the cohorts with lower $\alpha$ to gain more confidence on the effect estimations. 

With a finalized cohorts set and the cohort-level effect estimations, we ran stochastic optimization analyses using both SGD and Adagrad options \footnote{SGD and Adagrad are different choices of the proximal projection operator in Algorithm \ref{algo:mcsa}. For further details, please see Section \ref{sec:reproducibility}}. We observe that the SGD method converges faster than Adagrad in the Notification use case. Figure \ref{fig:prophetsp} shows the convergence of metric and constraint for the SGD use case over 100,000 iterations. Offline analysis suggested that by optimally choosing the caps based on a member's network and notifications quality features, we can lift the sessions metric while satisfying the other constraint metrics (e.g., Notification CTR, Total Disables). 

\subsection{Online Experiments}
\label{sec:online-experiments}

Online A/B tests have become the gold standard in evaluating new product developments in web-facing companies. We ran A/B tests to further validate our method, and report results comparing three variants against the existing baseline of a fixed global cap in Table \ref{tab:online_capnotification}. 
The first variant is the personalized cap treatment learnt using the proposed approach. Due to the high cost of implementing the pipeline and launching online A/B tests, we implemented only the cohort-level solution $CT.ST$ among the earlier proposed methods. 
The second and third variant exposes members to different caps based on a heuristic cohort definition. We grouped members into four segments according to their visit frequency on LinkedIn: members who visit daily, weekly, monthly, and dormant. We then manually selected the suitable cap parameter for each cohort to balance between the impacts on success and constraints metrics. Heuristic Cap A applies slightly larger caps compared to Heuristic Cap B which leads to higher impacts on Sessions but largely violating the constraints on Notification Sends and Notification Total Disables.

All reported metric lifts in Table \ref{tab:online_capnotification}) have a $p$-value of less than 0.05, otherwise we report the lift as neutral since it is not statistically significant. As can be seen from the results, our personalized cap treatment showed significant positive impact on Sessions, while the impact on the constraint metrics remained within acceptable bounds. Our method also outperforms the both heuristic solutions in terms of achieving the best gains on Sessions metric and balancing the negative impacts on the constraint metrics.

\begin{table}[ht]
\centering
\begin{tabular}{p{2.6cm}|p{1.5cm}|p{1.5cm}|p{1.5cm}} 
\hline\hline 
Metrics & ATE \% & ATE \% & ATE \%\\ 
& Personalized Cap &  Heuristic Cap A & Heuristic Cap B\\
[0.5ex]
\hline 
Sessions &  $\boldsymbol{+1.39\%}$  & $+1.31\%$ & $+0.54\%$ \\
Notification Sends & $ \boldsymbol{+1.64\%}$ & $+6.62\%$ & $+3.07\%$ \\
Notification CTR & $-1.24\%$ & $-1.73\%$ & -1.18\% \\
Total Disables & Neutral & $+9.23\%$ & Neutral\\
\hline 
\end{tabular}
\caption{Notification Cap Experiment Results} 
\label{tab:online_capnotification}
\end{table}

\section{Related Work}
\label{sec:relatedwork}
Estimating heterogeneity in causal effect is an important research area and in recent years, more researchers are applying supervised machine learning methods to the problem~\cite{uplift, Foster2011,imai2013, taddy2014}. Many researchers apply tree models and corresponding ensemble methods since they can automatically identify nonlinear relationships and interactions ~\cite{green2012, Athey7353,stefan2015forest}. In this body of work, the Causal Tree and Forest algorithms from \cite{Athey7353} and \cite{stefan2015forest} stands out as a popular choice which introduced the novel ``honest estimation'' concept and showed substantial improvements in coverage of confidence intervals. 
\cite{pmlr-v104-du19a} also explored similar ideas in leveraging heterogeneous effects estimation models (includes Causal Tree and Forest algorithms) in improving user retention.

In this paper, we have also described an algorithm for stochastic optimization with multiple expectation constraints based on the CSA algorithm developed in \cite{lan2016algorithms}. Traditionally, stochastic optimization routines were solved either via sample average approximation (SAA) \cite{wang2008sample, kleywegt2002sample} 
or via stochastic approximation \cite{robbins1951stochastic}.  
For more details on the above, please refer to \cite{benveniste2012adaptive} and \cite{spall2005introduction}. 
Very recently \cite{yu2017online} developed a method of stochastic online optimization with general stochastic constraints by generalizing Zinkevich's online convex optimization. Their approach is much more general and as a result, maybe not the optimal approach to solving this specific problem. For more related works in stochastic optimization literature, we refer to \cite{lan2016algorithms} and \cite{yu2017online} and the references therein.

\section{Discussion}
\label{sec:discussion}
In this paper, we propose a set of novel algorithms to personalize treatment selection using causal heterogeneity as estimated from randomized experiment data. The methods all consist of two major stages: estimating heterogeneous effects at cohort or member level; and solving an optimization problem to identify the optimal treatment variant for each member. We conducted simulation studies to evaluate and compare the proposed methods together with the global allocation baseline at different noise levels. We then applied the cohort-level estimation paired with the stochastic optimization method ($CT.ST$) on the LinkedIn Notifications system for personalizing daily notification caps. Both offline analysis and online experiments show strong results that validate our approach and the system has been fully deployed at Linkedin. In addition, the solutions produced by the approach can be applied in any online system to personalize treatment selection of any kind with minor overhead in terms of memory and latency.

Despite the previous focus of causal heterogeneity estimation, our work is the first complete proposal of exploiting or leveraging such heterogeneity to improve member experience and deliver bigger business wins. An important side effect of personalized treatment selection is its suitability in building more inclusive experiences as our proposed methods allow for different treatment selections for under-represented groups especially when these groups of members differ from the average population in their response to a treatment.
A few non-trivial, but likely impactful extensions for future consideration include: 
\begin{enumerate}
\item The cost of collecting online A/B test data could be large in terms of time and potential negative member experiences for some use cases. Designing a more cost-efficient data collection framework or leveraging observational data to achieve the same performance would be beneficial. 
\item Users can potentially move in and out of cohorts. Extending this framework to incorporate the dynamic nature of cohorts could be an interesting research topic.  
\item As mentioned in Section \ref{sec:causal}, future work on generating one single optimal cohort definition based on effects from multiple treatments with various metrics of interests could further improve the method. 
\end{enumerate}

\bibliographystyle{ACM-Reference-Format}
\bibliography{prophet} 
\pagebreak
\section{Reproducibility}
\label{sec:reproducibility}

In this section, we describe some further details that should help the reader to reproduce the methodology described in this paper. We also provide the \texttt{R} code that has been used to run simulation studies and all experiments. Due to the sensitive nature of the data, we are not able to disclose that actual data used in our experiments but the practitioner can use the details in this section and the code provided to run this system for their dataset.

\subsection{Details on Algorithm \ref{algo:mcsa}}
The proximal projection function $P_{\xb_t}(\cdot)$ used in Algorithm \ref{algo:mcsa} is defined as
\begin{align}
\label{eq:prox_proj}
P_{\xb_t}(\cdot) := \argmin_{z \in \Xcal} \left\{ \langle \cdot, z \rangle + B_{\psi_t}(z,\xb_t) \right\}
\end{align}
where $B_{\psi_t}(z,\xb_t)$ is the Bregman divergence with respect to the 1-strongly convex function $\psi_t$, defined as
\begin{align}
\label{eq:bregman}
B_{\psi_t}(x, y) = \psi_t(x) - \psi_t(y) - \langle \nabla \psi_t(y), x - y \rangle.
\end{align}
For more details we refer the reader to \cite{duchi2011adaptive}. In this paper, we have focused on two specific forms of functions $\psi_t$.
\begin{enumerate}
\item \textit{Stochastic Gradient Descent}: In this case, we use $$\psi_t(x) = x^Tx.$$
\item \textit{Adagrad}: Here we pick $ \psi_t(x) = x^T H_t x $ where $$H_t = \delta \Ib + diag\bigg( \sum_{j=1}^t h_j h_j^T\bigg) ^{1/2} $$
and $h_j$ are the chosen gradients in steps 5 and 8 of Algorithm \ref{algo:mcsa}.
\end{enumerate}
The comparison of each of the above two choices has already been discussed in Section \ref{sec:experiments}. Since we could not attach the real data due to to privacy concerns, we are adding a simulated result, which the reader can reproduce using the code given below. 

\begin{figure}[!h]
  \centering
  \includegraphics[width = 0.6\linewidth]{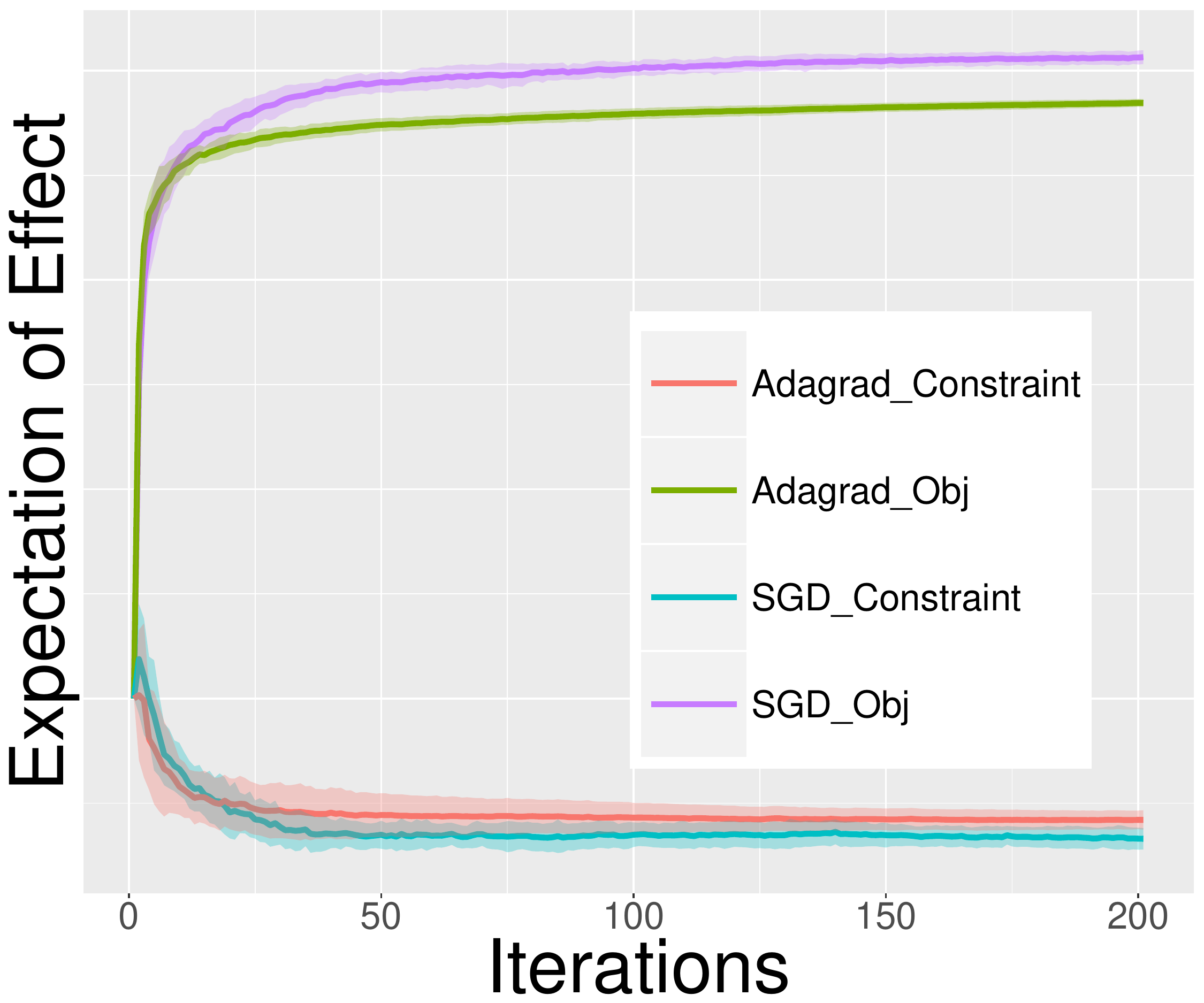}
  \caption{Simulated Results}
  \label{fig:simulated}
\end{figure}

We consider a problem having a success metric and two guardrail metrics and a treatment having 10 different levels. We generate a random distribution for each metric $k$ and each treatment level $j$. We run the algorithm for $N = 200$ iterations and choose $50$ samples to estimate the constraint at each step. We repeat both SGD and Adagrad 10 times and show the growth of the objective and the restriction on the constraint as the iterations increase in Figure \ref{fig:simulated}.

Note that we are not claiming that one method is better than the other. However in our simulations, we have usually observed that in simpler problems, SDG tends to converge faster, while in tougher problem, Adagrad tends to work better.  

\subsection{Code Details}
We share example scripts for conduct simulation analysis in examining the proposed methods and stochastic optimization algorithms in the following Github link: \url{https://github.com/tuye0305/prophet}. The source code contains five main files:

\begin{enumerate}
\item \texttt{multipleRunSimulation.R} - This code can generate multiple runs of simulation study flow (including generate simulation data, apply all proposed approaches to calculate optimal policy $\xb^*$ and evaluate the solution using simulated test datasets).
\item \texttt{generateSimulationDataUtils.R} - This file have the util functions associated with generating simulation data with self-initiated causal DAG.
\item \texttt{cohortLevelProphetUtils.R} - This file includes the functions to run cohort-level solution paired with stochastic optimization ($HT.ST$, $CT.ST$).
\item \texttt{mergeTreeUtils.R} - This file covers the util functions for merging trees trained with multiple metrics and treatment definitions which was described in Algorithm  \ref{algo:mergeTree}.
\item \texttt{memberLevelProphetUtils.R} - This file covers the util functions to enable the member-level estimations (Causal Forest and Two-Model method with Random Forest models) paired with deterministic optimization ($CF.DT$, $TM.DT$).
\item \texttt{StochasticOpSimulation.R} - This code runs the stochastic optimization routine to identify the optimal parameter in each cohort. Running the code as is should generate the plot as shown in Figure \ref{fig:simulated}.
\end{enumerate}

We have used the open source \texttt{R} libraries simcausal \cite{simcausal} to generate simulation datasets. We then applied the causalTree \cite{ctR}, randomForest \cite{rfR} and Generalized Random Forests \cite{grf} libraries to identify the cohorts or estimate effects for each treatment $j$ and metric $k$. Using these, the entire methodology discussed in this paper can be easily reproduced for any similar problem of interest for small scale problems.
\end{document}